# Review of Cost Reduction Methods in Photoacoustic Computed Tomography

**Afreen Fatima[1,2¥], Karl Kratkiewicz[1¥], Rayyan Manwar[1¥], Mohsin Zafar[1], Ruiying Zhang[3], Bin Huang[4], Neda Dadashzadesh[5], Jun Xia[6] and Mohammad Avanaki[1,7,8*]**

[1]*Department of Biomedical Engineering, Wayne State University, Detroit, MI, USA*

[2]*Department of Electrical & Computer Engineering, Wayne State University, Detroit, MI, USA*

[3]*Edan Medical, USA, Sunnyvale, USA*

[4]*3339 Northwest Ave, Bellingham, WA, USA*

[5]*R&D Fiber-Laser department, Coherent Inc., East Granby, CT, USA*

[6]*Department of Biomedical Engineering, The State University of New York, Buffalo, NY, USA*

[7]*Department of Neurology, Wayne State University School of Medicine, Detroit, MI, USA*

[8]*Molecular Imaging Program, Barbara Ann Karmanos Cancer Institute, Wayne State University, Detroit, MI, USA*

[¥] These authors have contributed equally.

**Running Head**:

Manuscript word count: 5723

*Corresponding Author: Mohammad Avanaki
Phone: +1 313 577 0703
E-mail: mrn.avanaki@wayne.edu

# Review of Cost Reduction Methods in Photoacoustic Computed Tomography

## Abstract

Photoacoustic Computed Tomography (PACT) is a major configuration of photoacoustic imaging, a hybrid noninvasive modality for both functional and molecular imaging. PACT has rapidly gained importance in the field of biomedical imaging due to superior performance as compared to conventional optical imaging counterparts. However, the overall cost of developing a PACT system is one of the challenges towards clinical translation of this novel technique. The cost of a typical commercial PACT system originates from optical source, ultrasound detector, and data acquisition unit. With growing applications of photoacoustic imaging, there is a tremendous demand towards reducing its cost. In this review article, we have discussed various approaches to reduce the overall cost of a PACT system, and provided a cost estimation to build a low-cost PACT system.

# 1. Introduction

Photoacoustic computed tomography (PACT) is a major configuration of photoacoustic imaging (PAI), a novel hybrid imaging modality that combines optical excitation of the target sample with acoustic detection that is generated due to thermal expansion of the sample as shown in Fig. 1. In PACT, high energy pulsed laser light is diffused to create a full field illumination that covers the sample (tissue), and photoacoustic waves are generated [1-3]. The waves around the tissue are collected by wideband ultrasound transducers [4, 5]. The detection scheme can be realized either by a single ultrasound transducer which rotates around the sample (Fig. 2a), a linear array (Fig. 2b), or a stationary ring array of 128, 256, or more number of transducer elements (Fig. 2c) [6-12]. The ultrasound waves collected from the object are acquired using a data acquisition (DAQ) unit, and an image reconstruction algorithm is used to reconstruct a PACT image. The image shows vascular and functional information of the tissue.

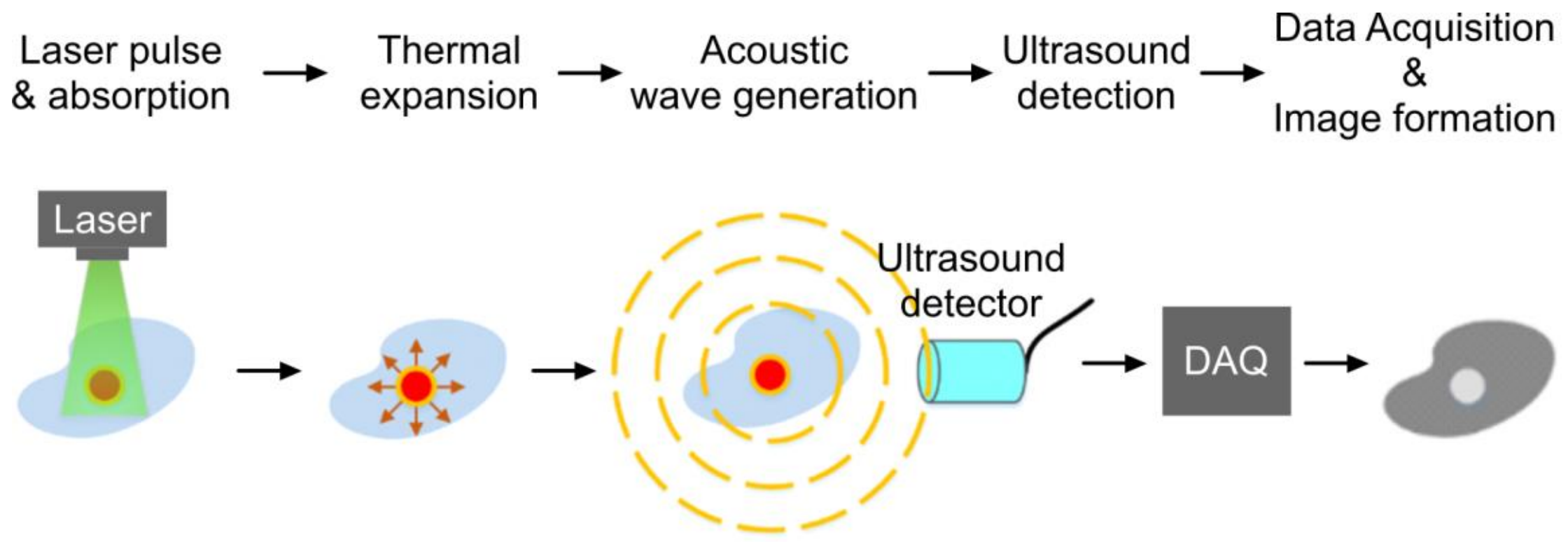

**Figure 1.** Principle of photoacoustic signal generation, detection, and acquisition.

The rising number of investments in research and development coupled with the constant pace of technological advancement leading to the development of hybrid imaging systems is primarily driving the growth of the global preclinical imaging market [13]. The global optical imaging market is anticipated to be approximately $1.9 billion by end of 2018, growing at a compound

annual rate (CAGR) of 11.37%. Among the optical imaging modalities, photoacoustic and near-infrared spectroscopy contributes ~6.85% combined [14]. PACT has gained even more popularity for deep tissue imaging (within the range of several centimeters [15-17]) where coarse resolution is acceptable. Over the past few years, various preclinical and clinical applications of this technique have been demonstrated, including functional brain imaging , small-animal whole-body imaging , breast cancer screening [1], and guidance of lymph node biopsy [18]. Several improvements have been applied to PACT to overcome its limitations [3, 19-26]. The main reason for the popularity of PAI compared to optical imaging methods is that, the acoustic scattering in tissue is about three orders of magnitude lower than optical scattering [3, 27-30].

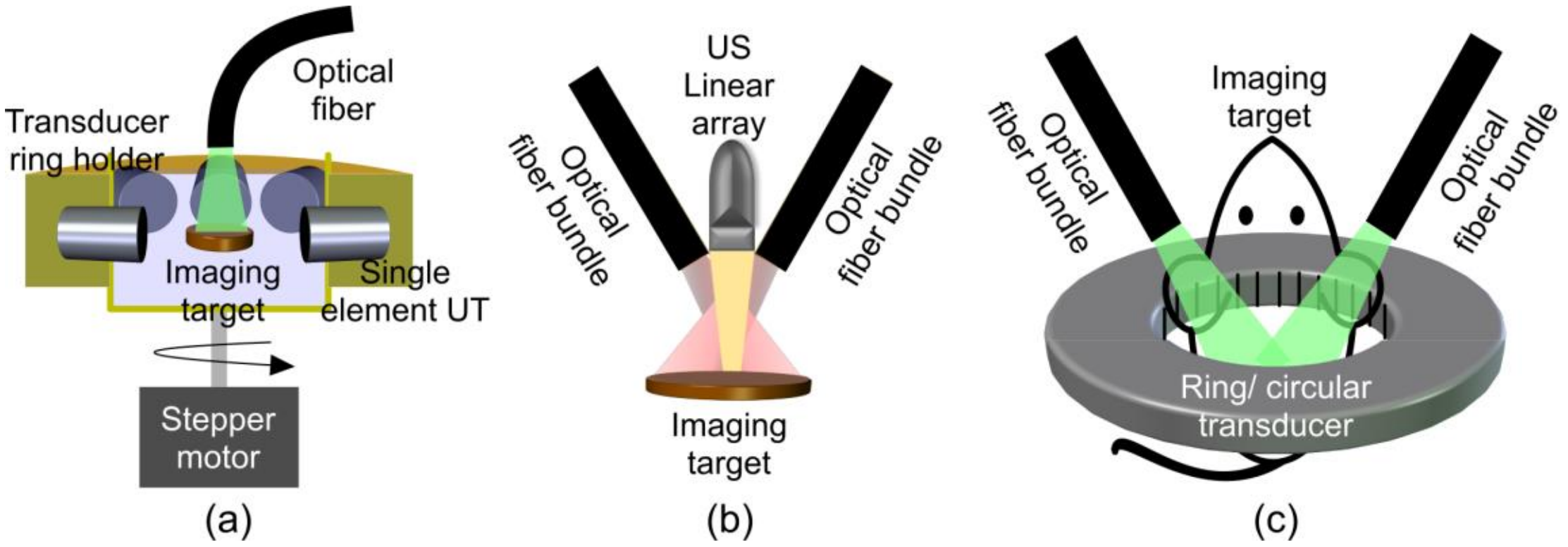

**Figure 2.** Typical configurations of a PACT. (a) Single/multi element transducers in ring fashion, (b) linear-array, and (c) ring/ circular transducer array.

Considering many advancements of photoacoustic imaging in PACT system development, it is still in its early stage for clinical use compare to established imaging modalities such as magnetic resonance imaging, positron emission tomography and computed tomography. The major hurdle is cost. The cost of PACT systems which are commercially available is provided in Table 1. The majority of the cost originates from three main components: (1) optical source, (2) ultrasound detector, and (3) DAQ unit.

Different optical sources for the implementation of a low-cost PACT system have been discussed in [31, 32]. Typically, a Q-switched Nd: YAG pulsed laser is employed in commercially available PACT systems that costs in the range of $15~$100K USD depending on the level of energy/pulse and the pulse width. Some of these laser sources provide built-in fixed wavelength options, typically at 532nm and 1064nm. There is an additional cost for continuous wavelength tuning feature (650-950nm) that is provided by optical parametric oscillator (OPO) and typically used for PA imaging of specific endogenous or exogenous chromophores. Commercial PACT systems come with application specific ultrasound transducer probe which are typically charged in the range of $1K~$200K. Based on the application and system configuration, the geometry (i.e. linear [33], two-dimensional (2D) [34], arc, ring [35], or hemisphere [36]), number of elements (64~2048), and center frequency (1~55MHz) of the transducer varies. Linear array-based configurations offer fast image acquisition; however, they rely on expensive hardware and data acquisition unit. The cost of PACT systems increases due to expensive software packages for system control, image reconstruction, display and post-processing (~$1K-$50K). The cost of a commercial PACT system is ~$50K-$500K. Vevo LAZR and LAZR X (Fujifilm VISUALSONICS, Canada) [37], Nexus 128+ (Endra Life Sciences, USA) [38], LOUISA 3D (TomoWave, Inc, USA) [39, 40], and MSOT inVision 128 (iTheraMedical GmbH, Germany) [33] are commercial PACT systems that provide real-time volumetric vascular/functional images [41-43].

**Table 1.** Typical price list of commercially available PACT systems.

| PACT system | Vevo LAZR X | Vevo LAZR | PAFT | MSOT inVision 128 | Nexus 128+ | LOUISA 3D |
|---|---|---|---|---|---|---|
| **Company** | Fujifilm VISUALSONICS Inc. ON, Canada | Fujifilm VISUALSONICS Inc. ON, Canada | PST Inc. TX, USA | iTheraMedical GmbH, Germany | Endra Life Sciences, MI, USA | TomoWave, Inc. TX, USA |
| **Approximate Cost** | ~950K | ~750K | ~315K | ~470K | ~375K | ~215K |
| **Image** | 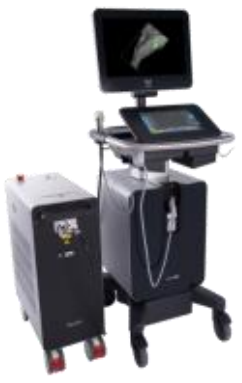 | 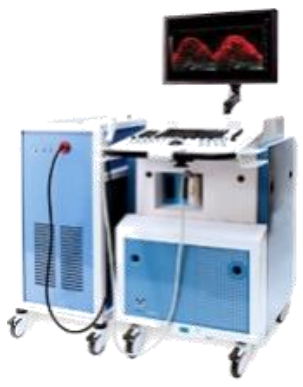 | 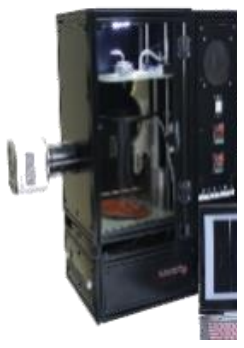 | 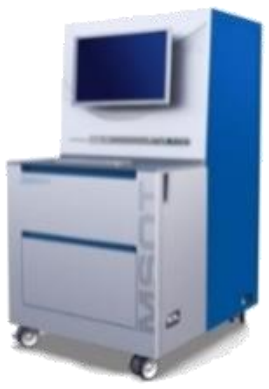 | 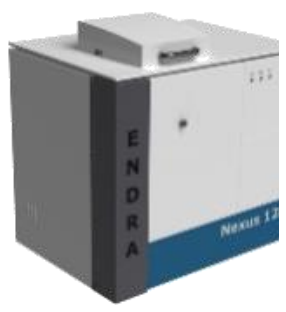 | 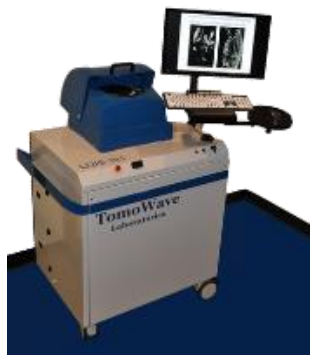 |
| **Application** | Oncology, cardiology, molecular and neuro biology | Same as LAZR-X | Small animal imaging | Real-time whole body imaging | Molecular, Tumor hypoxia etc. | Breast cancer research |

Due to the increasing number of applications of PACTs, there is an urgent need to make cost-effective, compact and portable PACT systems. In the following sections, we have discussed the cost-reduction methods in PACT systems. Firstly, we reviewed alternate affordable laser sources, followed by in-depth discussion of low-cost ultrasound transducer configurations, and finally examples of low-cost DAQs. Based on the elaborative review on the low-cost solutions, at the end we have estimated the cost of a low cost PACT system.

## 2. Conventional cost-reduction methods

Here, we have discussed the conventional approaches for cost reduction of PACT systems in terms of their major hardware components. In section 2.1, we have reviewed the light sources. Low-cost ultrasound detection apparatus for the PACT systems are discussed in section 2.2. Finally, DAQs are reviewed in section 2.3.

### 2.1. Affordable compact light sources

Although typically in PACT systems, nanosecond pulsed laser sources with average pulse energy of 10~100mJ are desirable [44], other, lower-cost light sources such as laser diodes (LD), light-emitting diodes (LED), flash lamps, electrically-pumped lasers, and optically-pumped lasers can be used [45-48]. Since the beam divergence and beam quality of the light sources used in the configuration of PACT are not required to be stringent, there is more options than expensive lasers with fine linewidth. In the following sub-sections, we review the light sources that are being used for low-cost PACT systems.

### 2.1.1. Laser diodes (LDs)

LDs are examples of electrically-pumped lasers. In LDs, the gain is obtained by an electrical current flowing through p-i-n structure of the semiconductor medium. LDs can be manufactured based on different technologies and wide variety of them are commercially available in laser industry [49-52]. LDs usually operate in the wavelength range between 750 nm to 980 nm [53, 54] and are also available in visible wavelengths ranging from 400 nm to 650 nm, where blood absorption is strong (>10 $cm^{-1}$) and water absorption is weak (<$10^{-3}cm^{-1}$) [53-55]. Attempts have been made to use LDs in handheld devices because of their compact size [56-60]. One of the LDs that has been developed to use in PACT system is developed by Kohl et al. [61] and Canal et al. [62]. This LD is a pulsed ultra-compact multi-wavelength LD array source which delivers pulse energies comparable to nanosecond Nd: YAG lasers. The laser system has a footprint of 20 $cm^2$ and provides pulses with a pulse-repetition rate (PRR) of 10 kHz, pulse energy of 1.7 mJ, and pulse width of 40 ns at single wavelengths of 808 nm, 915 nm, 940 nm, or 980 nm. Daoudi, et al. [58] also developed an LD that is appropriate for PACT system. This LD is made based on diode stacks technology [58]. In this laser, the diodes are driven by a customized laser driver (Brightloop, France). This LD is measured to produce pulses with pulse width of 130 ns, PRR of 10 kHz, and emission wavelength of 805 nm. The PACT system that uses this laser has experimentally been utilized for phantoms and *in-vivo* imaging of human proximal interphalangeal joint as shown in Fig. 3 [58]. There are several other studies that have used light source array to increase the laser energy [56, 63]. The only limitations of using array light source is the large beam size and circularity issues of the light source in addition to the requirement of using a larger pump driver [59, 64]. This limitation is especially problematic in photoacoustic microscopy where a tight focus is needed, not in PACT where a diffused light is used. If the laser energy and pulse width of the LD array is the same as a high energy Nd:YAG laser, we expect to see the same quality PA images.

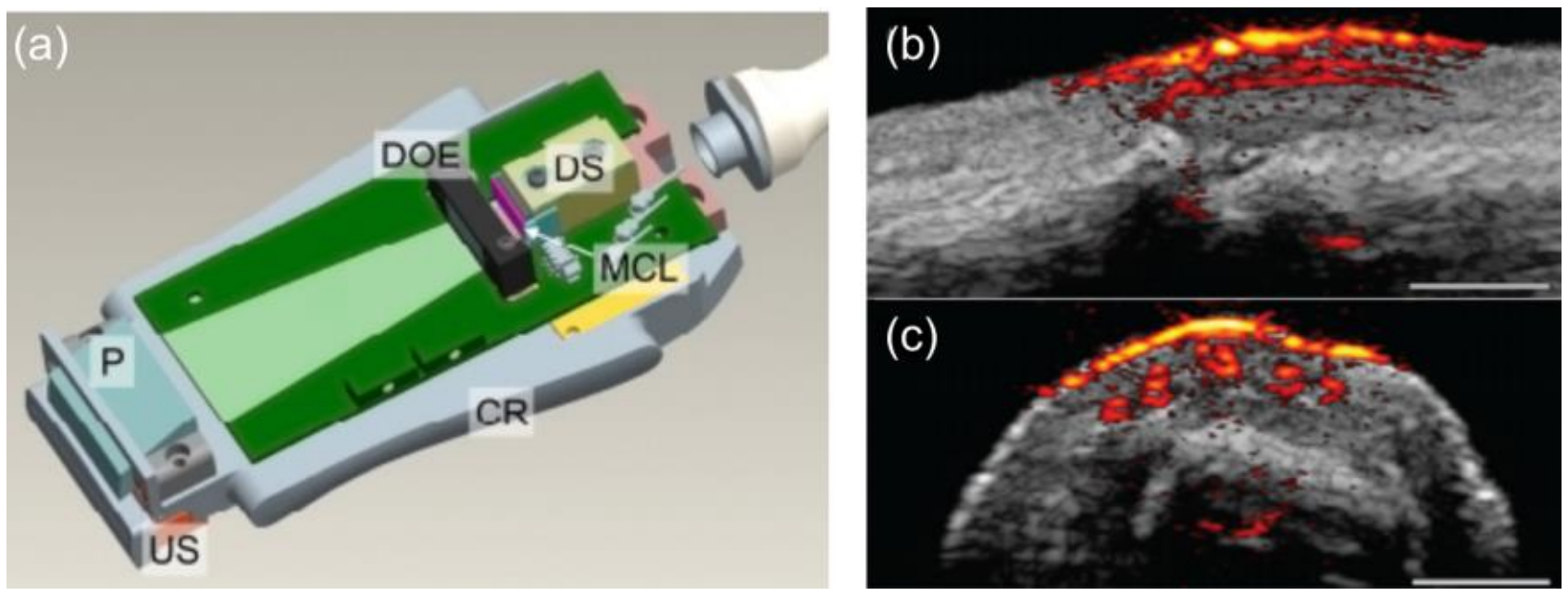

**Figure 3.** (a) A schematic of the handheld PACT probe. US: ultrasound array transducer, P: deflecting prism, DOE: diffractive optical elements, DS: diode stack, MCL: micro-cylindrical lenses, CR: Aluminum cooling rim. Photoacoustic/ultrasound images of a human proximal interphalangeal joint in (b) sagittal and (c) transverse planes. Reproduced with permission from [58].

## 2.1.2. Light Emitting Diodes (LEDs)

LED is a category of light sources that has been used in PACT system due to their extremely compact size, lightweight, long lifetime, and low-cost potentials. Although the fundamental process of light generation in LEDs is similar to LDs, LEDs do not generate stimulated emission [65]. Therefore, their optical spectrum is broad bandwidth, their spatial coherence is low, and they are not technically considered as laser sources. However, they are still proper light sources for applications that do not require a narrow linewidth light source such as PA imaging. LEDs can be manufactured in a wide wavelength range [66-68]. Broad wavelength range of LEDs makes them ideal candidates for spectroscopic applications in PA field [69, 70]. The enormous progress in creating high-power LEDs has brought attentions to them as an efficient low-cost light source for large output signals. LEDs provide pulse energies of tens of μJ when operated in pulsed mode at low duty cycles (<1%) and are driven by ten times of their rated current [69]. PreXionLED (PreXion Corp.,Tokyo, Japan), is LED arrays that can provide pulses with 1 kHz PRR, 100 ns pulse width, and 200 μJ pulse energy at wavelengths of 750 nm, 850 nm, and 930 nm [68, 70]. Using LED arrays, imaging of PA contrast agent, i.e., Indocyanine green (ICG) under several centimeters of chicken breast tissue was successfully demonstrated [67]. Additionally in [70], these LED arrays have been utilized in various clinical applications such as assessment of peripheral microvascular function and dynamic changes, diagnosis of inflammatory arthritis, and detection of head and neck cancer. The LED arrays (PreXion Corporation), and a resulting image from the PACT setup is shown in Fig. 4. Kang, *et. al*. also used LED arrays for tube phantoms imaging on mouse skull and human skull samples [71]. LED arrays was also used by Dai, *et. al.* for performing *in vivo* mouse ear vasculature imaging.

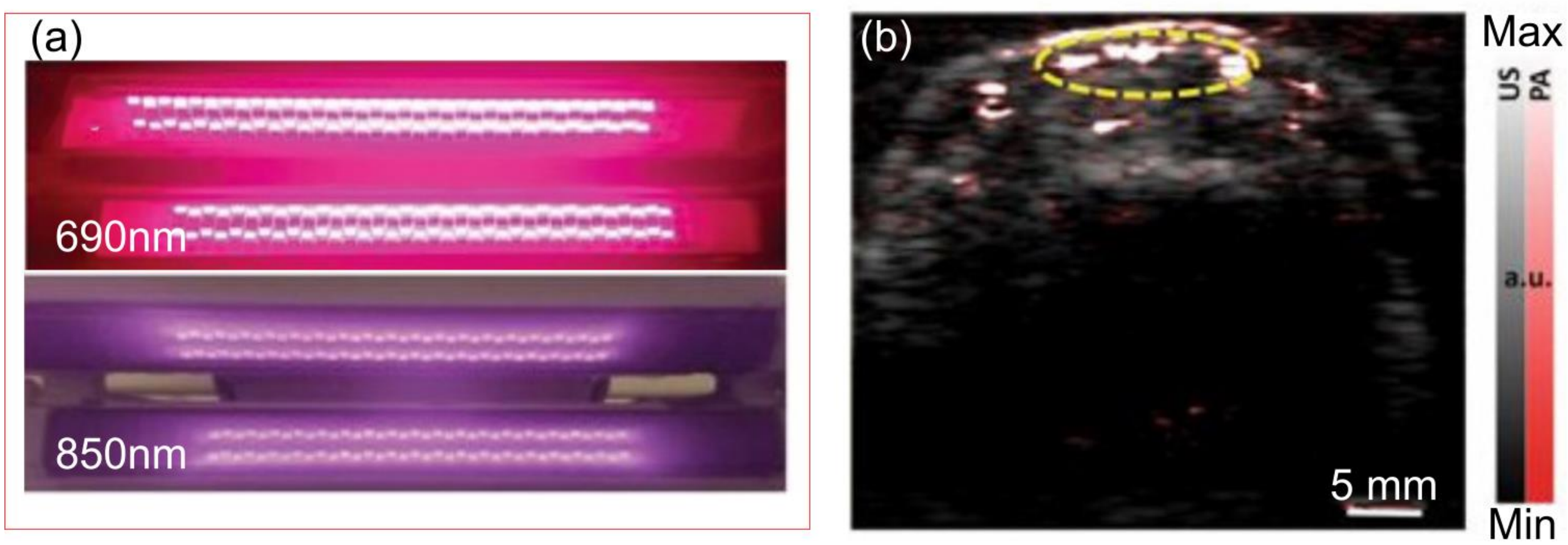

**Figure 4.** (a) Photo of a pair of dual-wavelength LED bars that emits 690-nm and 850-nm light alternatively. (b) PA (pseudo color) and US (gray scale) combined image showing the microvessels in the cross-section of a human finger. Reproduced with permission from [70].

### 2.1.3. Xenon Flash Lamp (XFL)

XFL is a very low cost high energy light source for PACT when single-mode operation and high spatial resolution are not required. XFLs have cathode and anode electrodes facing each other in a Xenon-filled glass bulb and they emit by arc discharge. XFLs emit in a broad spectrum from UV to IR, and they have high intensity, high stability, and long life-time. Pulsed XFLs with high pulse energy and microsecond pulse widths have been used as a low-cost alternative light source in PACT systems. Wong, et. al. demonstrated *in vivo* rat imaging using an XFL. The vasculature in a rat body was clearly observed in the PACT image as shown in Fig. 5 [72]. Since the optical illumination from XFLs meets requirement of human's laser exposure, XFLs can potentially be applied to human tissue for imaging purposes.

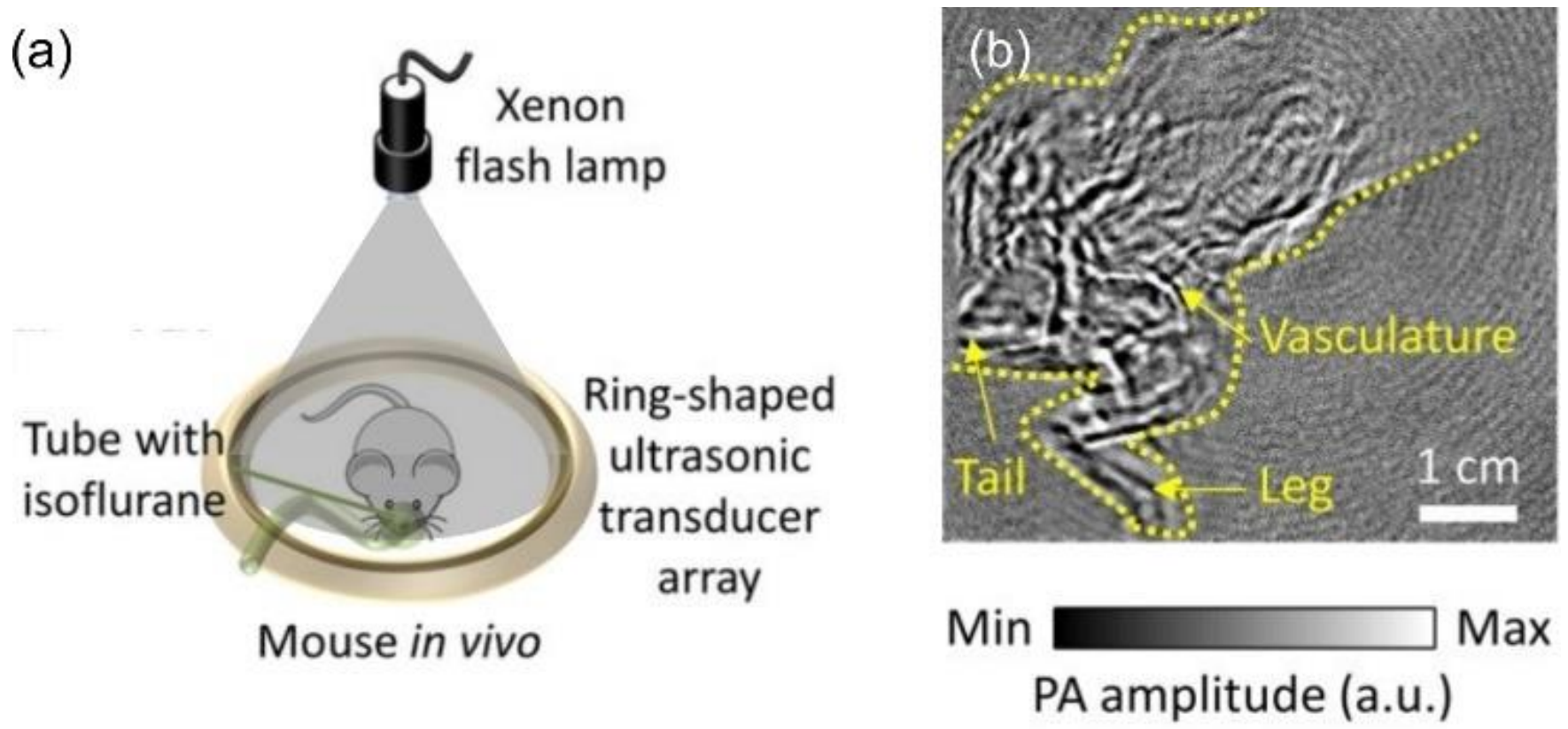

**Figure 5.** (a) Xenon flash lamp illumination, (b) PACT image of a mouse body *in vivo*. Reproduced with permission from [72].

### 2.1.4. Solid-state Diode-Pumped Laser (DPL)

Main categories of optically pumped lasers are lamp-pumped lasers (LPL) and diode-pumped lasers. Solid-state DPLs are known as either bulk lasers, that have a crystal as their active media, or fiber lasers, in which light propagates in active optical fibers as the gain media [73-75]. Solid-state DPLs can be designed to produce power levels of a few mWs to multiple KWs due to their high energy conversion efficiency.  Solid-state DPLs make compact lasers and a better candidate for PA imaging than bulky water-cooled lamp-pumped lasers. Wang, *et. al.* used a compact high-power DPL (Montfort Laser GmbH Inc., Germany) for deep tissue *in vivo* PACT imaging [76]. It has a miniature size of $13.2 \times 14.0 \times 6.5$ cm$^3$, a weight of 1.6 kg, average power output of 4W with a high pulse energy of up to 80 mJ at the wavelength 1064 nm with a PRR up to 50 Hz (see Fig. 6a). Using this laser system as shown in Fig. 6b, they successfully imaged murine whole-body vascular structures and cardiac functions *in vivo*, and mapped the arm, palm and breast vasculatures of living human subjects.

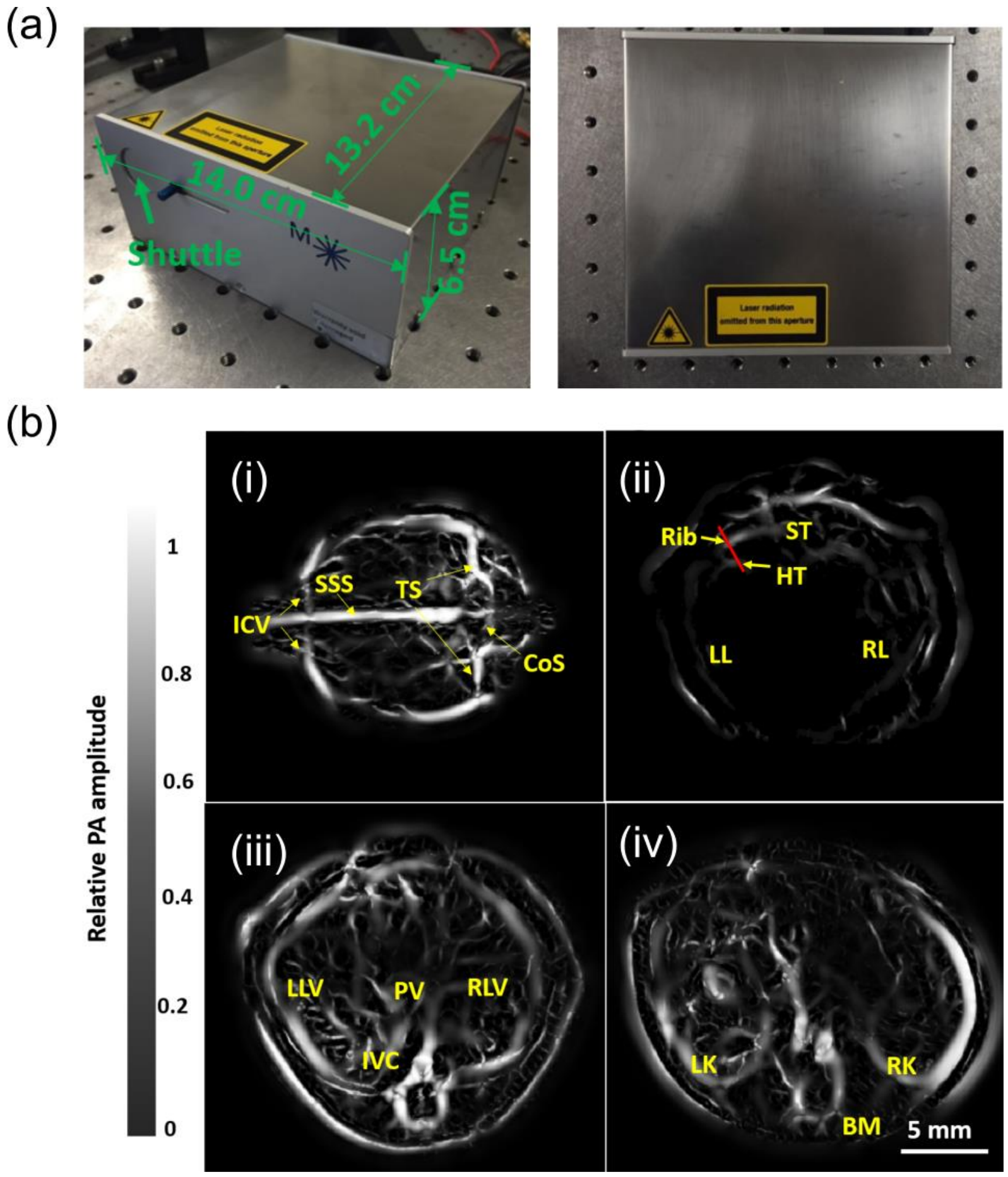

**Figure 6.** (a). Photographs of a diode-pumped Nd:YAG laser; (b) mouse anatomical and functional imaging; (i) PA image of cerebral vasculature of a mouse brain. CoS: confluence of sinuses, ICV: inferior cerebral vein, SSS: superior sagittal sinus, TS: transverse sinus, (ii) PA cross sectional image of the heart region. ST: sternum, HT: heart, LL: left lung, RR: right lung, (iii) PA cross sectional image of liver region. LLV: left lobe of liver, PV: portal vein, RLV: right lobe of liver, IVC: inferior vena cava, (iv) PA cross sectional image of kidney region. Reproduced with permission from [76].

## 2.1.5. Continuous Wave (CW) lasers

Continuous-wave (CW) operation mode of a laser stands for the case that the laser is continuously pumped and continuously emits light. The CW emission of a laser can occur in either a single resonator mode or in multiple modes. Some of the CW lasers that are inexpensive, compact, and durable with an average power of several Watts, are generated by LDs [44]. LeBoulluec, *et. al*. designed a frequency domain- PA imaging system and used a CW LD as the light source with central wavelength of 785 nm and 100 mW output power [77]. They tested the imaging system on several tissue-like phantoms, where, CW PA detection used a narrow-band ultrasonic transducer and a lock-in amplifier with high sensitivity and strong noise rejection. In [78], another inexpensive, compact and durable single frequency intensity modulated CW laser diode based PA system was reported as shown in Fig. 7 (a). This system was successfully employed to image ~3mm deep vessels in tissue (Fig. 7b).

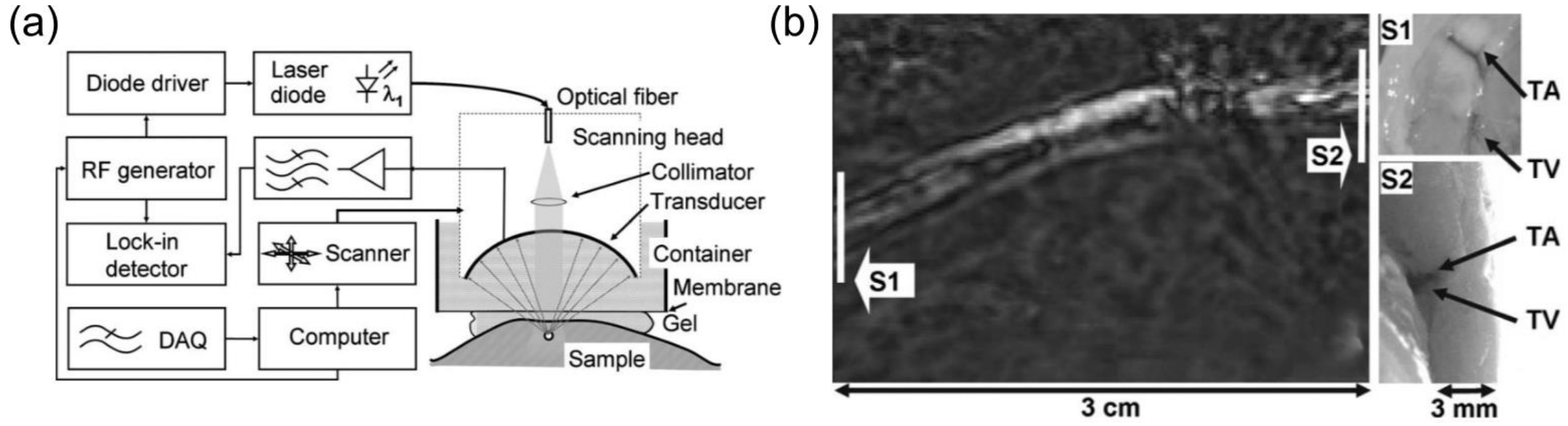

**Figure 7.** (a) Schematic of the CW PACT system, (b) amplitude image of a rabbit tibial artery (TA) and a tibial vein (TV) acquired with the CW PA imaging system. Reproduced with permission from [79]

In Table 2, we have summarized different laser sources that have been used in PACT imaging systems with their specifications and particular imaging applications.

**Table 2.** Summary of alternative illumination sources for PACT

| Type | Repetition rate (kHz) | Wavelength (nm) | Type/Model | Application |
|---|---|---|---|---|
| LD | 10 | B | Quantel, Paris, France | Hand-held *in vivo* of human finger joint [80] |
| | 7 | B | Quantel DQQ1910-SA-TEC | *In vivo* mouse brain imaging [81] |
| | 30 | B, C | NA | Characterization of laser source [57] |
| **LED** | 0.5 | B | SST-90, Luminus, USA | Tissue mimicking phantom [69] |
| | 40 | A | EDC405-1100, Marubeni | *In-vivo* mapping of mouse ear [82] |
| | 4 | B, C | HDHP LED | *Ex-vivo* transcranial [71] |
| | 1, 4 | A, B | PreXion Corporation | Blood oxygenation, [70] |
| **DPL** | Up to 0.5 | C | Montfort Laser GmbH M-NANO-Nd: YAG-10ns-50-INDR_PR135 | *In-vivo* deep tissue imaging [76] |
| **XFL** | 10-100 | A, B, C | L4634, Hamamatsu Photonics K. K. | Phantoms and small animal [72] |
| **CW*** | NA | B | GH0781JA2C, Sharp | Rabbit ear vasculature [79] |
| | NA | B | L785P100, Thorlabs | [77] |
| | NA | C | IPG Photonics, MA | PVC phantom [78] |

*Only CW laser energy is provided in Watts
Optical window: A <600nm, 600nm< B <900nm, and C >900nm

## 2.2. Less expensive photoacoustic signal detection methods

Another major pricy component of a PAI system is the detection unit. A considerable cost reduction is achieved if we can utilize the existing ultrasound detection system and integrate the PAI system into it. Typically, piezoelectric thin films are used to fabricate ultrasound transducers based on piezoelectric effect [83]. The most popular piezoelectric materials are the polycrystalline ferroelectric ceramic materials, such as barium titanate (BaTiO3) and lead zirconate titanate (PZT), which consist of randomly oriented crystallites (grains), separated by grain boundaries [84]. These materials are much less expensive than single crystals (lead magnesium niobate–lead titanate (PMN–PT) and lead zinc niobate–lead titanate (PZN–PT)) but offer strong piezoelectric properties along polarization axes. Piezoelectric polymers such as polyvinylidene (PVDF) and its copolymer with trifluoroethylene (TrFE) have also been found to be useful for producing high frequency transducers [85, 86]. PZT has been the dominant material for the active elements of transducers and arrays [87-90]. A summary of the low-cost PA ultrasound detection methods is provided in Table 3.

### 2.2.1. Single element transducers

One way to reduce the overall cost of a PACT system is by reducing the number of transducers. Several strategies have been evaluated. The scanning of a single-element finite-size transducer, due to its simplicity and high sensitivity, is widely used in the implementation of PACT [91, 92]. The disadvantage of this configuration is a low temporal resolution. Scanning a number of single element transducers, also called multi element transducer PACT (MET-PACT), can greatly reduce the cost for a PACT system while preserving a high temporal resolution. Several groups including ours have demonstrated the use of this configuration with a high frame rate (~ 1-2 seconds) and several-centimeter penetration depth [93-97]. The major drawback in using multiple single element transducers in a PACT is that all of them cannot be placed experimentally at the same distance from the scanning center. Each of them rotates around the sample in concentric circles with slight difference in radius (~ 1-3 mm). Based on trial and error, this issue can be resolved. However it is time consuming and becomes even more complex when the number of transducers increases. In [98], a MET-PACT with 16 single-element 5MHz ultrasound transducers (Technisonic, ISL-0504-GP) have been implemented. More can be read in [98, 99]. The transducers spatially separated with 22.5° from each other were inserted and fitted along the circumference of the circular ring made up of polyactic acid (PLA) plastic (15cm diameter) (see Fig. 8). A data correction algorithm was performed to resolve the difficulty in placing the transducers at the same distance from the scanning center. The use of partial view detection with reflectors is another method to decrease the numbers of ultrasound transducers needed [98, 100].

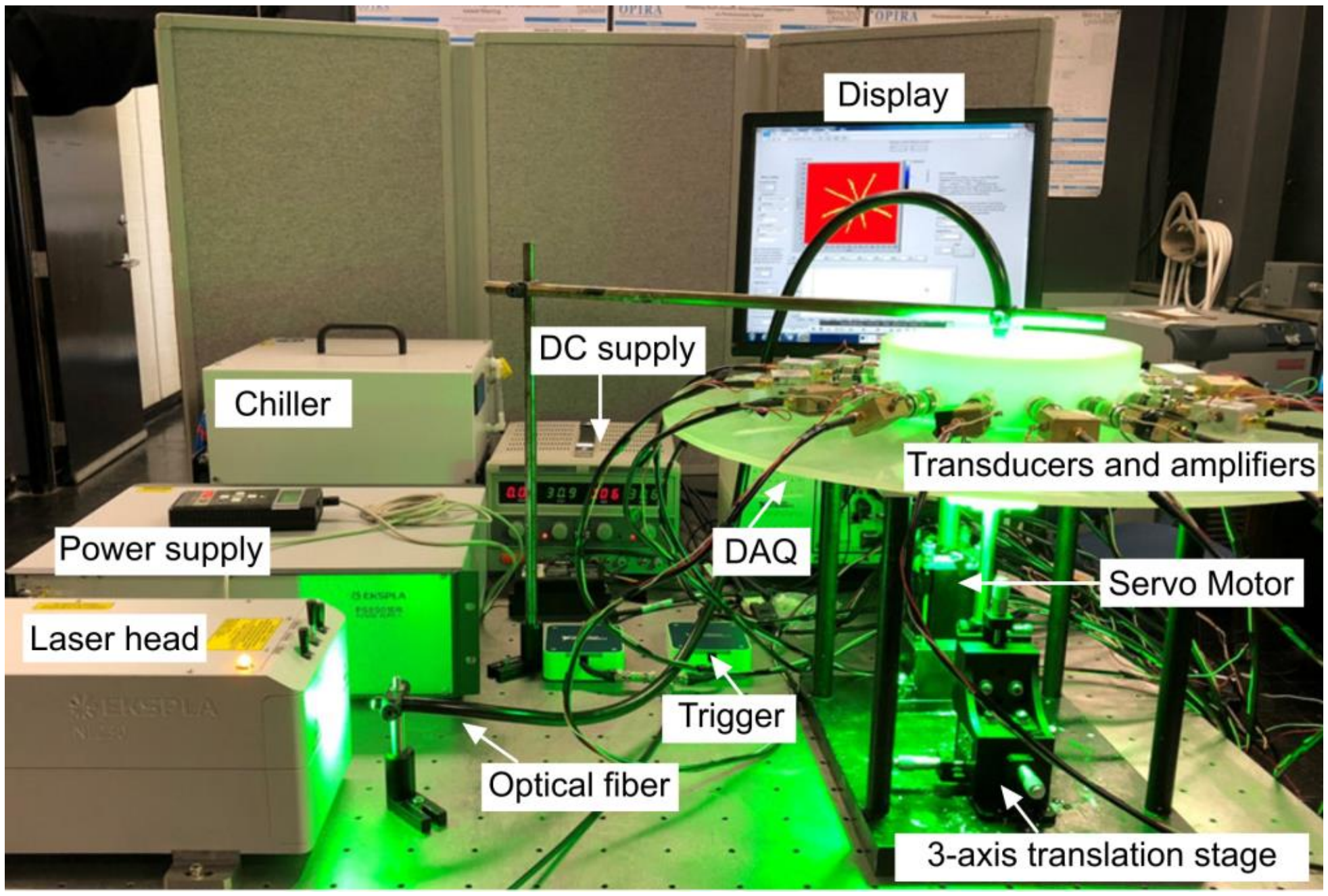

**Figure 8.** A multi element transducer PACT system comprised of a portable EKSPLA laser head, power supply for the laser, chiller, DC supply for the motor driver, NI DAQ, NI trigger, servo motor, motor gear, 3-axis translation stage, and 16 sets of transducers and amplifiers. Reproduced with permission from [98].

### 2.2.2. Linear Array transducers

In addition to MET-PACT, linear detector arrays have also been studied. A large planar detector can be a piezoelectric plate much larger than the object to be imaged [101]. Linear arrays can be easily manufactured and fabricated in batches, hence the production cost is lower and yield-rate can be very high as compared to custom-made curved or ring arrays [102, 103]. Most importantly, these arrays have been commonly used in clinical applications and can easily be integrated with the light sources. Four linear array US/PA transducers are shown in Fig. 9 where they are coupled with different configurations of light sources.

Linear array-based photoacoustic imaging could be the closest configuration of photoacoustic imaging to the clinic. The appearance of the PA probe in this configuration is similar to the US probe that clinicians are used to work with, in addition to the hand-held and compact nature of them.

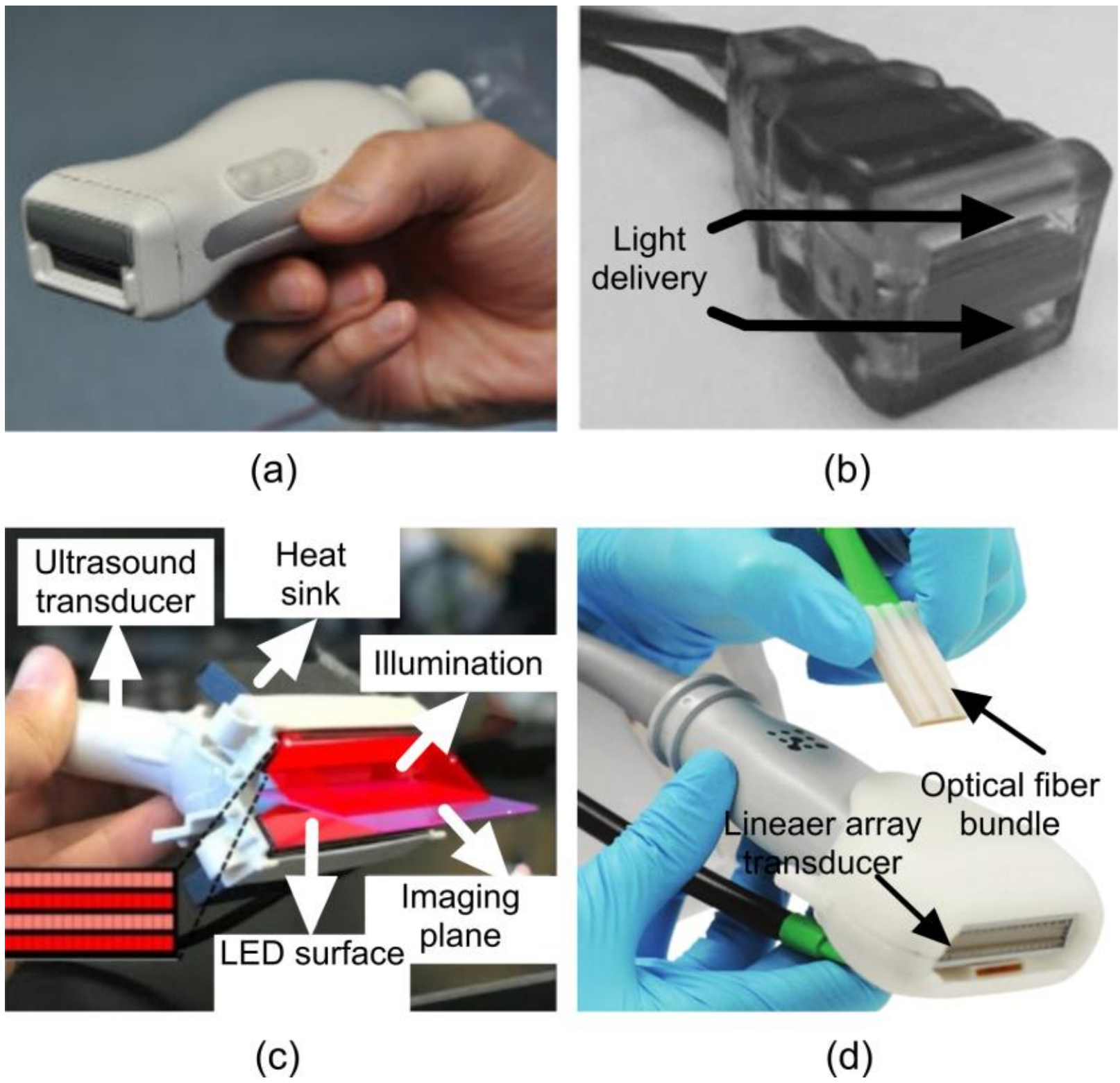

**Figure 9.** Linear array transducers used in photoacoustic signal detection, (a) Reproduced with permission from [104], (b) Reproduced with permission from [18], (c) Reproduced with permission from [67], and (d) Visualsonic PA probe. Reproduced with permission from Fujifilm VisualSonics Inc.

Several groups have implemneted linear array PACT. In [58, 104], the ultrasound detection is performed with an ultrasound pulse/receiver array (based on the commercial Esaote SL3323 ultrasound probe) composed of 128 elements, each with a length of 5 mm and a pitch of 0.245 mm. The array has a central frequency of 7.5 MHz and a measured −6 dB bandwidth of around 100%. Li et al.  developed and engineered a handheld probe-based photoacoustic imaging system that was built on an open-platform ultrasound imaging system with a 128-element linear array ultrasound scanner (7 MHz, LA0303, S-sharp, Taiwan). Several real-time systems for small animal experiment have also been reported . Due to the limited aperture of linear array transducers, features with high aspect-ratio or orientations oblique to the transducer surface suffer distortion and azimuthal resolution is reduced. Chinni et al. have designed and implemented a novel acoustic lens based focusing technology that improves the overall aperture of the linear array transducer [105]. In this system, all generated PA waves from the laser get focused simultaneously by the lens onto an image plane to produce a 2D mapping of the absorber distribution. The lens is not digital however, equally effective focusing can be achieved by judicious choice of lens material for signals over a large band of frequencies. A linear array transducer was used to capture the focused PA signals. By scanning the linear array in the image plane, both C-scan and volumetric data sets were generated. The acoustic lens did not consume electrical power, eliminates the dedicated software and hardware for beamforming and image reconstruction. The system is compact and economical with lower system design and fabrication cost.

**Table 3.** Summary of less expensive PA signal detection methods

| Type | Central Frequency (MHz) | Bandwidth (%) | No. of elements | Model/ Company | Ref |
|---|---|---|---|---|---|
| Single element/ multi element PACT | 2.25, 5 | 70 | 2 | V323-SU, V309-SU, Olympus NDT | [4] |
| | 3.5 | 88 | 1 | V383, Panametrics | [16] |
| | 10 | 100 | 1 | XMS-310, Panametrics | [106] |
| | 5 | 75 | 16 | ISL-0504-GP, Technisonic | [98] |
| Linear array PACT | 7.5 | 100 | 128 | SL3323, Esaote | [58, 104] |
| | 7 | 85 | 128 | LA0303, S-sharp. | [107] |
| | 10 | 80.9 | 128 | PreXion Corporation | [67] |
| | 5 | 60 | 128 | L7-4, ATL/Philips | [76] |
| | 5 | NA | 320 | EZU-PL22, Hitachi | [108] |

## 2.3. Low-cost DAQ and other associated electronic components

Eliminating the need for a high-speed sampling rate requirement may be the most efficient way to reduce the size and cost of data acquisition (DAQ) [109]. Gao et al. developed a palm-size sensor to use a rectifier circuit (Fig 10a) to convert the high-frequency PA signal to low-frequency [109]. The proposed photoacoustic receiver could potentially reduce the cost and device size efficiently. The photoacoustic signal detected by an oscilloscope and its rectified DC signal detected by the proposed palm-size sensor after normalization were in good agreement. The ultrasound transducer used in this case was 1 MHz with 60% fractional bandwidth (V303-SU, Olympus, USA). The rectified photoacoustic DC signal is then sampled by a low-speed analog to digital converter (ADC) with 10 kHz sampling rate and processed by a low-cost microprocessor board (e.g., Arduino Uno).

Low-cost computer-based DAQ has also been investigated. Mihailo et al. used a sound card as DAQ for a low-cost, portable photoacoustic instrument [110]. The device consists of a detection unit comprising of a photoacoustic cell with an embedded laser diode, a photodiode, an electret microphone ($60 \times 40 \times 40$ mm$^3$), and a signal processing and power supply unit in a box containing batteries and electronics ($160 \times 140 \times 60$ mm$^3$) (Fig. 10b). A PC with an Asus SiS7012 16-bit sound card is required to operate the device and for data processing. The sound card generates the signal for the laser diode or LED modulation and processes signals from the microphone and photodiode. The experimental points are obtained at a maximum of 48 kHz sampling rate. The described portable photoacoustic apparatus showed good agreement with the results of conventional photoacoustic devices.

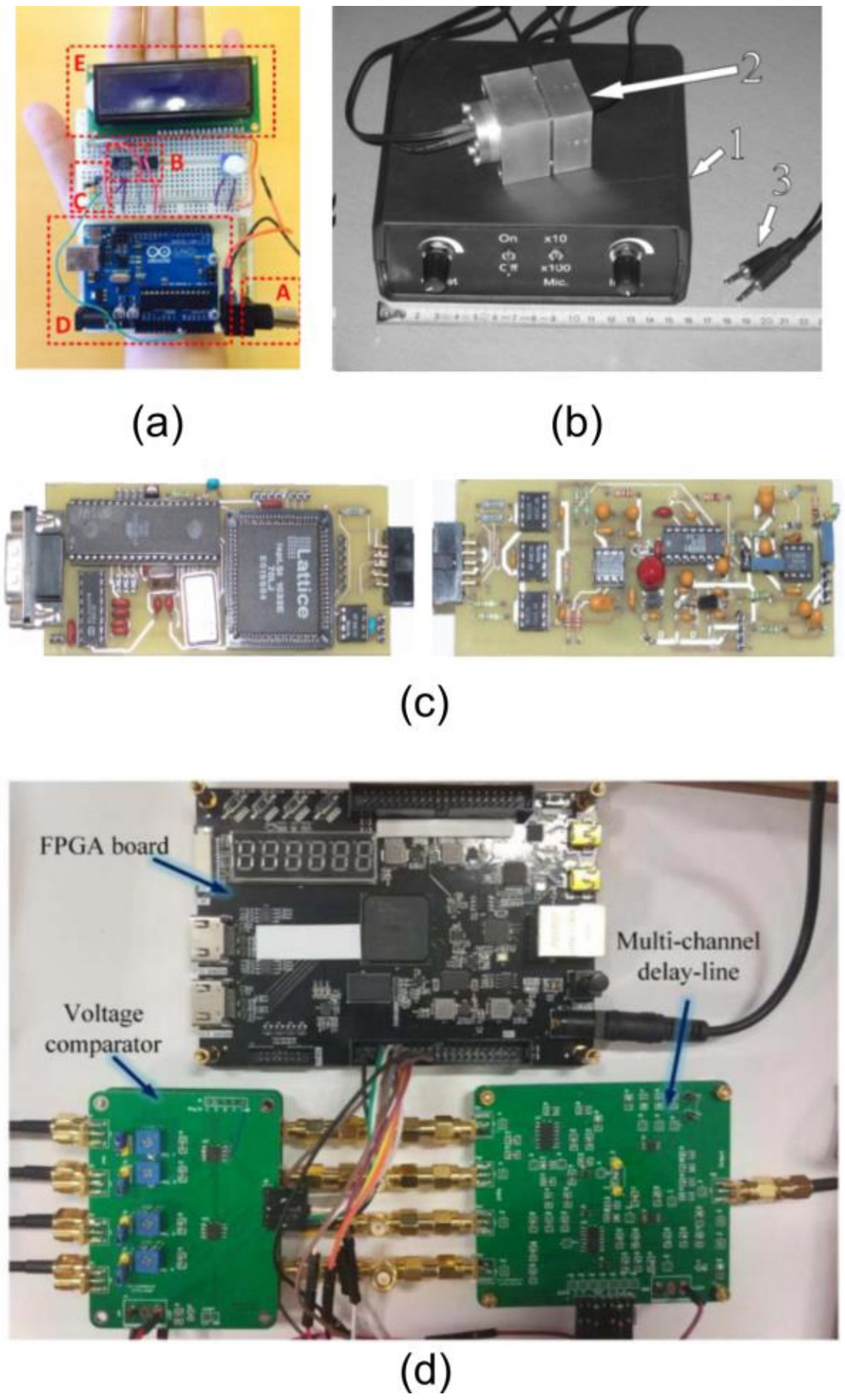

**Figure 10.** (a) Photograph of the proposed photoacoustic sensor that can fit in the palm of one's hand. Reproduced with permission from [109], A: PA signal input, B: amplifier, C: rectifier, D: Arduino board, and E: display, (b) photoacoustic cell. Its mass is 1.70 kg. (1) The box which contains the electronics and the batteries, (2) photoacoustic cell with an embedded microphone, a laser diode and a photodiode, (3) standard 3.5 mm stereo connectors. Reproduced with permission from [110], (c) photograph of the miniature DAQ and control dedicated for PA measurements. Reproduced with permission from [111], (d) a photograph of the delay-line module. Reproduced with permission from [112].

A miniature system for data acquisition and control, dedicated for photoacoustic measurements, was developed by Starecki and Grajda [111]. Based on the concept of virtual instruments (intensive use of electronics and digital signal processing), most of the expensive and/or inconvenient elements (mechanical chopper or lock-in amplifier) were eliminated. This system is comprised of two parts. First, the analog module that is used for two-stage PA signal amplification and conversion to digital samples. Second, the digital module acts as a controller of the analog part and simultaneously as a slave for external master controller. This device (Fig. 10c) is capable of data acquisition and basic digital signal processing (i.e. averaging). The advantages of the designed device are very low cost (< $100), small size, and high functional flexibility. Most of the measurement factors, e.g. analog path gain, light source modulation frequency, number of samples per period, etc. are programmable and can be changed at run-time.

In [112], a multiway delay line module was proposed to reduce the numbers of DAQ channels required for PACT system and hence the overall cost. The circuitry consisted of multiple voltage comparator (MVC), precise rectify circuit (PRC), delay line module (DLM), multiway adder (MA), and an FPGA board (Fig. 10d). The DLM (4:1) module was connected immediately to the PA detection array. It accommodated three delay units and a transmission unit, which combined the four inputs into one output series in time. The first PA signal was utilized as a reference signal and used to reconstruct the PA signals later. Additionally, the other three PA signals were transformed by three delay line units with different delay time. Finally, four synchronized input signals were combined into one pulse train output in time series. The feasibility of the module was tested through imaging phantoms and the reconstructed images were comparable to the conventional detection method. To improve the image quality, the authors suggested to use denoising technique as a part of the post-processing method.

Implementation of real-time signal sampling was made possible using 8-bit RISC Atmel microcontroller as shown in Fig. 10c. It has an internal memory of 512 bytes and 10-bit ADC which is sufficient for 250ksps data acquisition and real-time operations like averaging. Due to in-circuit programmable components, upgrades of the device were easy. Small size and low power consumption made the device a very good choice for implementation of low-cost photoacoustic instruments.

Another method to implement the data acquisition part of the PACT is by using very high sampling frequency DAQs and multiplexing hardware/software. For instance, utilizing a 500 MHZ DAQ for an application where 50 MHZ sampling frequency is required, along with a hardware or software multiplexing technique, can produce 10 channel of 50 MHz data. We have summarized different low-cost DAQs and associated electronic components in Table 4.

**Table 4.** Summary of low cost DAQs and associated electronic components

| Ref | Components | Model/ Company | Price (USD) | Potential application |
|---|---|---|---|---|
| | Low speed ADC, 10kHz | Arduino Uno | 20~50 | Oxygen saturation and temperature monitoring |
| | Low-noise amplifiers | LT 1226, Linear Technology | 10~20 | |
| | Sound card, 16-bit, 48kHz | SiS7012, ASUS | 40~200 | *In-vivo* and *in-situ* plant leaves |
| | Low noise amplifiers | OP37, Analog Devices | 5~10 | |
| | Microcontroller, 10-bit ADC, 250kHz | AT90S8535, Atmel | 15~30 | Water vapour measurement |
| | MVC, PRC, DLM, MA, and FPGA | Xilinx Inc. | 50~300 | Imaging of wire phantom |

MVC: multiple voltage comparator, PRC: precise rectify circuit, DLM: delay line module, MA: multiway adder, FPGA: field programmable gate array

## 3. Estimated cost for developing a low-cost PAI system

Various alternative options have been discussed to develop a low-cost PACT system. In the following, we suggest a low-cost PACT system. Based on the review on different alternative low-cost options, LD arrays would be an optimal choice considering the delivered optical energy. An LD array with the central wavelength of 905 nm (905D1S03X, Laser Component Co., Bedford, New Hampshire), a peak power of 6 W, and pulse width of 55 ns operating at maximum repetition rate of 20 kHz was used in [59]. Stacked configuration will provide maximum output of 650W with a pulse length of 150ns. For PA signal detection, multiple single element transducers with a stepper motor scanning system could provide 360° full view of the target with relatively fast acquisition. Utilizing a multi-channel DAQ with a very high sampling frequency and multiplexing techniques could be one low cost acquisition method [112]. Instead of image reconstruction software that is typically developed and customized for a specific PACT system, one can use an open source freeware such as k-wave toolbox or MATLAB (~300 USD). Estimated cost of individual component is provided in Table 5.

**Table 5.** Cost estimation for a low-cost PACT system

| Component | Type | Price (USD) | Company |
|---|---|---|---|
| **Laser** | PLD arrays, 20 kHz, 55ns pulse width | 10~15K | Laser Component Co., Bedford, New Hampshire |
| **Optical fiber bundle and optics** | Dia: 1 mm, NA: 0.25, Diffuser: 120 grit | ~500 | Newport Co., USA |
| **US transducer** | 8 single elements, 2-10 MHz | 2~5K | Olympus IMS |
| | 128-channel linear arrays, 2-10 MHz | 10~15K | Sonic Concepts, WA, USA |
| **Amplifiers** | Low-noise 40dB | ~300 | Mini Circuits RF/Microwave Components, USA |
| **DAQ (single/ multi-channel)** | 8 channel differential, 200KS/s | ~500 | NI-PCI 6034E, National Instruments, USA |
| | 128 channels, 40 MS/s (Linear array) | ~30K | PST Inc., USA |
| **Servo motor** | 24-70 VDC | ~400 | Applied Motion Products Inc. |
| **Software** | MATLAB | ~300 | MathWorks, USA |

Low-cost PACT systems will have limited applications where, (1) a high temporal resolution is required, (2) a tunable laser is needed for spectroscopy purposes, (3) or a very high energy is needed for deep tissue imaging.

## 4. Summary and Conclusions

PAI is a high contrast, non-invasive and non-ionizing imaging modality with extensive applications in anatomical and functional imaging that has been used for diagnostics purposes and cancer studies [59, 99, 113-118]. Cost is one major factor hindering PACT's implementation in clinical field. This paper described a comprehensive review on studies and progresses on ways to lower the cost of a PACT system. At first, light sources, as one major component that is significantly contributing to the cost of PACT, were explained. Various light sources such as LDs, LEDs, XFL, and DPL were explained. The type of the light source in the configuration of a PACT is chosen based on the compactness, temporal resolution, energy and beam profile requirements of the imaging application. These low-cost light sources reduce the overall cost of a PACT system, which eventually leads to easy clinical translatability, however, they come with limitations such as low SNR PA signals which consequently degrade the quality of the reconstructed images. We also mentioned that to attain sufficient SNR, LD or LED arrays are the optimum solution. We further explained that in applications where low power consumption is preferred, semiconductor LDs are the best choice. We then discussed different PA signal detection techniques. We mentioned that instead of using complex and expensive ultrasound transducer arrays, commercially available multiple single-element transducers or linear arrays can be utilized to reduce the cost. The limitation of these transducer arrays is where mechanically moving the transducers is not feasible or a high temporal resolution is required (> 20Hz). Additionally, we described methods working based on acoustic reflection to improve the limited view of the PA detection system. We discussed different hardware and software based alternatives to implement low cost DAQs. We described how using very high sampling frequency DAQs and multiplexing hardware/software we can implement low-cost data acquisition systems. The role of wavefront shaping and signal postprocessing in low-cost PACT systems is essential to improve the image quality [5, 22, 119-124]. For example, selection of an optimum laser pulse duration /pulse shape profile has greatly improved the resolution of the reconstructed image [32, 125]. According to [60], it was shown that in deep tissue imaging, utilizing a laser with an optimum pulse duration, the frequency-dependent acoustic attenuation was greatly reduced. Advanced beamforming techniques has also been studied; for example, Bell *et. al*, showed that the PA image contrast of imaging targets at depths 5-15 mm was improved by 11-17 dB with short-lag spatial coherence (SLSC) beamforming when compared to conventional delay-and-sum (DAS) beamforming [120, 121, 126]. Additionally, adaptive denoising technique in [122, 127] indicated that the proposed method increased the SNR of PA signals with fewer acquisitions as compared to common averaging techniques. Another effective technique that has been prescribed to improve the quality of PACT images is compressed sensing and sparse acquisition [128-131]. Other improvement methods are studied in [5, 118, 122]. By implementing these algorithms on multi-core processors such as digital signal processors (DSPs) [132], utilizing parallel processing methods [133], the temporal resolution of data acquisition will not be affected. Based on the review of different approaches, we have estimated that to develop a low-cost PACT system for shallow depth imaging applications with comparable performance to the commercial ones, one should spend ~$20K-$80K (see Table 5); the exact cost depends on the specification of the PACT system. By the advancement of the light sources, reducing their size and increasing their energy [134], new and low-cost ultrasound detectors with extremely small element size and low power consumption, e.g., micromachined piezo-electric/ capacitive ultrasound transducers (P/CMUTs) [135-137], and development of ultrahigh speed microcontroller-based DSPs with a large memory and number of bits, the cost of PACT systems will gradually be reduced while the quality of their images will be improved. Finally, utilizing all optical ultrasound detection methods [138-140], will solve the limitation of PA imaging probes where they require ultrasound conducing material between the imaging sample and the probe, and thus widen the applicability of the PACT devices.

Low-cost PACT systems facilitate the emergence of novel, compact, rapid *in vivo* imaging systems to enhance the accessibility and popularity of PAI in clinical applications. We hope this comprehensive review article will help the researchers recognize the different aspects of developing an affordable yet high-performance PACT system for both research purpose and clinical applications.

## Author contribution

**Afreen Fatima:** Writing – original draft, and Writing – review & editing **Karl Kratkiewicz:** Formal analysis, Investigation, Visualization, Writing – original draft, and Writing – review & editing. **Rayyan Manwar:** Data curation, Formal analysis, Investigation, Project administration, Software, Validation, Visualization, Writing- Original draft, and Writing - -review & editing. **Mohsin Zafar:** Formal analysis, Investigation, and Writing – review & editing. **Ruiying Zhang**: Investigation, Methodology, Validation, and Writing- Original draft. **Bin Huang:** Methodology. **Neda Dadashzadesh**: Data curation, Validation, and Writing- review and editing. **Jun Xia:** Data curation, Methodology, Supervision, Validation, Writing- Original draft, and Writing - -review & editing. **Mohammad Avanaki:** Conceptualization, Data curation, Formal analysis, Funding acquisition, Methodology, Project administration, Resources, Software, Supervision, Validation, Writing- Original draft, and Writing - -review & editing.

## Funding

This research was partially funded by the American Cancer Society, Research Grant number 14-238-04-IRG and the Albert and Goldye J. Nelson grant.

## Acknowledgement

We would like to thank Dr. Jalilian from University of Michigan, for her insightful discussion.

## Conflict of interest

The authors declare no conflict of interest.

## References

[1] Kruger, Robert A, et al., *Dedicated 3D photoacoustic breast imaging.* Medical physics, 2013. **40**(11).

[2] Lou, Yang, et al., *Impact of nonstationary optical illumination on image reconstruction in optoacoustic tomography.* JOSA A, 2016. **33**(12): p. 2333-2347.

[3] Wang, Lihong V and Hu, Song, *Photoacoustic tomography: in vivo imaging from organelles to organs.* science, 2012. **335**(6075): p. 1458-1462.

[4] Upputuri, Paul Kumar and Pramanik, Manojit, *Performance characterization of low-cost, high-speed, portable pulsed laser diode photoacoustic tomography (PLD-PAT) system.* Biomedical optics express, 2015. **6**(10): p. 4118-4129.

[5] Omidi, Parsa, et al., *A novel dictionary-based image reconstruction for photoacoustic computed tomography.* Applied Sciences, 2018. **8**(9): p. 1570.

[6] Gamelin, John Kenneth, et al., *Curved array photoacoustic tomographic system for small animal imaging.* Journal of biomedical optics, 2008. **13**(2): p. 024007.

[7] Xia, Jun, et al., *Whole-body ring-shaped confocal photoacoustic computed tomography of small animals in vivo.* Journal of biomedical optics, 2012. **17**(5): p. 050506.

[8] Xia, Jun, et al., *Three-dimensional photoacoustic tomography based on the focal-line concept.* Journal of biomedical optics, 2011. **16**(9): p. 090505.

[9] Deán-Ben, X Luís, et al., *Spiral volumetric optoacoustic tomography visualizes multi-scale dynamics in mice.* Light: Science & Applications, 2017. **6**(4): p. e16247.

[10] Li, Lei, et al., *Single-impulse panoramic photoacoustic computed tomography of small-animal whole-body dynamics at high spatiotemporal resolution.* Nature biomedical engineering, 2017. **1**(5): p. 0071.

[11] Merčep, Elena, et al., *Transmission–reflection optoacoustic ultrasound (TROPUS) computed tomography of small animals.* Light: Science & Applications, 2019. **8**(1): p. 18.

[12] Mohammadi, Leila, et al., *Skull's Photoacoustic Attenuation and Dispersion Modeling with Deterministic Ray-Tracing: Towards Real-Time Aberration Correction.* Sensors, 2019. **19**: p. 345.

[13] Research, Grand View, *Preclinical Imaging Market Analysis By Product Type (Devices: CT, MRI, PET/SPECT, Multi-modal, Optical, Ultrasound, Photoacoustic (PAT), Reagents and Services) And Segment Forecasts To 2024*, in *Preclinical Imaging Market Size, Industry Report, 2024*. 2016.

[14] Ltd, MarketsandMarkets Research Private, *Optical Imaging Market (2013-2018) – Technology Trends And Applications Of Optical Coherence Tomography (OCT), Hyper Spectral Imaging (HSI), Near Infrared Spectroscopy (NIRS) And Photo- Acoustic Tomography (PAT) In Clinical Diagnostics, Clinical Research And Life Sciences With*

*Market Landscape Analysis - Estimates Up To 2018*, D. Sokolov, Editor. 2013: marketsandmarkets.com. p. 33.
[15] Ku, Geng and Wang, Lihong V, *Deeply penetrating photoacoustic tomography in biological tissues enhanced with an optical contrast agent.* Optics letters, 2005. **30**(5): p. 507-509.
[16] Wang, Xueding, et al., *Noninvasive laser-induced photoacoustic tomography for structural and functional in vivo imaging of the brain.* Nature biotechnology, 2003. **21**(7): p. 803.
[17] Weber, Judith, Beard, Paul C, and Bohndiek, Sarah E, *Contrast agents for molecular photoacoustic imaging.* Nature methods, 2016. **13**(8): p. 639.
[18] Garcia-Uribe, Alejandro, et al., *Dual-modality photoacoustic and ultrasound imaging system for noninvasive sentinel lymph node detection in patients with breast cancer.* Scientific reports, 2015. **5**: p. 15748.
[19] Mohammadi-Nejad, Ali-Reza, et al., *Neonatal brain resting-state functional connectivity imaging modalities.* Photoacoustics, 2018.
[20] Mahmoodkalayeh, Sadreddin, et al., *Low Temperature-Mediated Enhancement of Photoacoustic Imaging Depth.* Scientific Reports, 2018. **8**(1): p. 4873.
[21] Mahmoodkalayeh, Sadreddin, et al. *Optimization of light illumination for photoacoustic computed tomography of human infant brain*. in *Photons Plus Ultrasound: Imaging and Sensing 2018*. 2018. International Society for Optics and Photonics.
[22] Meimani, Najme, et al., *A numerical analysis of a semi-dry coupling configuration in photoacoustic computed tomography for infant brain imaging.* Photoacoustics, 2017. **7**: p. 27-35.
[23] Wang, Lihong V, *Tutorial on photoacoustic microscopy and computed tomography.* IEEE Journal of Selected Topics in Quantum Electronics, 2008. **14**(1): p. 171-179.
[24] Xu, Minghua and Wang, Lihong V, *Photoacoustic imaging in biomedicine.* Review of scientific instruments, 2006. **77**(4): p. 041101.
[25] Hennen, Stella N, et al., *Photoacoustic tomography imaging and estimation of oxygen saturation of hemoglobin in ocular tissue of rabbits.* Experimental eye research, 2015. **138**: p. 153-158.
[26] Gamelin, John, et al., *A real-time photoacoustic tomography system for small animals.* Optics express, 2009. **17**(13): p. 10489-10498.
[27] Tavakolian, Pantea, *Potential for Photoacoustic Imaging of Neonatal Brain*. 2014, The University of Western Ontario.
[28] Panchal, Rakshita, et al., *Vibration analysis of healthy skin: toward a noninvasive skin diagnosis methodology.* Journal of biomedical optics, 2019. **24**(1): p. 015001.
[29] Turani, Zahra, et al., *Optical Radiomic Signatures Derived from Optical Coherence Tomography Images to Improve Identification of Melanoma.* Cancer Research, 2019: p. canres. 2791.2018.
[30] Zarei, Mehrdad, et al. *Simultaneous photoacoustic tomography guided diffuse optical tomography: a numerical study*. in *Photons Plus Ultrasound: Imaging and Sensing 2019*. 2019. International Society for Optics and Photonics.
[31] Yao, Qingkai, et al., *Low-cost photoacoustic imaging systems based on laser diode and light-emitting diode excitation.* Journal of Innovative Optical Health Sciences, 2017. **10**(04): p. 1730003.
[32] Zhong, Hongtao, et al., *Review of low-cost photoacoustic sensing and imaging based on laser diode and light-emitting diode.* Sensors, 2018. **18**(7): p. 2264.

[33] Medical, iThera. *MSOT inVision*. 2019 [cited 2019; Available from: https://www.ithera-medical.com/products/msot-invision/.
[34] Vaithilingam, Srikant, et al., *Three-dimensional photoacoustic imaging using a two-dimensional CMUT array.* IEEE transactions on ultrasonics, ferroelectrics, and frequency control, 2009. **56**(11): p. 2411-2419.
[35] Passler, K, et al. *Annular piezoelectric ring array for photoacoustic imaging*. in *European Conference on Biomedical Optics*. 2011. Optical Society of America.
[36] Lv, Jing, et al., *Hemispherical photoacoustic imaging of myocardial infarction: in vivo detection and monitoring.* European radiology, 2018. **28**(5): p. 2176-2183.
[37] Inc., FUJIFILM VisualSonics. *Vevo LAZR*. 2019 [cited 2019; Available from: https://www.visualsonics.com/product/imaging-systems/vevo-lazr.
[38] Inc., ENDRA. *ENDRA Life Sciences Adds Nexus 128+ to Their 3D Preclinical Photoacoustic CT Lineup*. 2016 [cited 2019; Available from: https://www.prnewswire.com/news-releases/endra-life-sciences-adds-nexus-128-to-their-3d-preclinical-photoacoustic-ct-lineup-300321461.html.
[39] Su, Richard, et al., *Laser optoacoustic tomography: towards new technology for biomedical diagnostics.* Nuclear Instruments and Methods in Physics Research Section A: Accelerators, Spectrometers, Detectors and Associated Equipment, 2013. **720**: p. 58-61.
[40] Tsyboulski, Dmitri A, et al., *Enabling in vivo measurements of nanoparticle concentrations with three-dimensional optoacoustic tomography.* Journal of biophotonics, 2014. **7**(8): p. 581-588.
[41] Bohndiek, Sarah E, et al., *Development and application of stable phantoms for the evaluation of photoacoustic imaging instruments.* PloS one, 2013. **8**(9): p. e75533.
[42] Levi, Jelena, Sathirachinda, Ataya, and Gambhir, Sanjiv S, *A high affinity, high stability photoacoustic agent for imaging gastrin releasing peptide receptor in prostate cancer.* Clinical Cancer Research, 2014: p. clincanres. 3405.2013.
[43] Needles, Andrew, et al., *Development and initial application of a fully integrated photoacoustic micro-ultrasound system.* IEEE transactions on ultrasonics, ferroelectrics, and frequency control, 2013. **60**(5): p. 888-897.
[44] Stylogiannis, Antonios, et al., *Continuous wave laser diodes enable fast optoacoustic imaging.* Photoacoustics, 2018. **9**: p. 31-38.
[45] Francis, TS and Yang, Xiangyang, *Introduction to optical engineering*. 1997: Cambridge University Press.
[46] Held, Gilbert, *Introduction to light emitting diode technology and applications*. 2016: Auerbach Publications.
[47] Shiraz, HG, *Principles of semiconductor laser diodes and amplifiers: analysis and transmission line laser modeling*. 2004: World Scientific.
[48] Träger, Frank, *Springer handbook of lasers and optics*. 2012: Springer Science & Business Media.
[49] Chow, Weng W and Koch, Stephan W, *Semiconductor-laser fundamentals: physics of the gain materials*. 2013: Springer Science & Business Media.
[50] Coldren, Larry A, Corzine, Scott W, and Mashanovitch, Milan L, *Diode lasers and photonic integrated circuits*. Vol. 218. 2012: John Wiley & Sons.
[51] Delfyett, Peter J, et al., *High-power ultrafast laser diodes.* IEEE Journal of Quantum Electronics, 1992. **28**(10): p. 2203-2219.

[52] Endriz, John G, et al., *High power diode laser arrays.* IEEE Journal of quantum electronics, 1992. **28**(4): p. 952-965.
[53] Thomas, R. L. , et al., *Subsurface flaw detection in metals by photoacoustic microscop.* J. Appl. Phys, 1980. **51**: p. 1152.
[54] Hoshimiya, Tsutomu, Endoh, Haruo, and Hiwatashi, Yoichiro, *Observation of surface defects using photoacoustic microscope and quantitative evaluation of the defect depth.* Japanese Journal of Applied Physics, 1996. **35**(5S): p. 2916.
[55] Jansen, E Duco, et al., *Temperature dependence of the absorption coefficient of water for midinfrared laser radiation.* Lasers in Surgery and Medicine, 1994. **14**(3): p. 258-268.
[56] Allen, Thomas J and Beard, Paul C, *Pulsed near-infrared laser diode excitation system for biomedical photoacoustic imaging.* Optics letters, 2006. **31**(23): p. 3462-3464.
[57] Canal, Celine, et al. *Short-pulse laser diode sources enabling handheld photoacoustic devices for deep tissue imaging*. in *European Conference on Biomedical Optics*. 2017. Optical Society of America.
[58] Daoudi, Khalid, et al., *Handheld probe integrating laser diode and ultrasound transducer array for ultrasound/photoacoustic dual modality imaging.* Optics express, 2014. **22**(21): p. 26365-26374.
[59] Hariri, Ali, et al., *Development of low-cost photoacoustic imaging systems using very low-energy pulsed laser diodes.* Journal of biomedical optics, 2017. **22**(7): p. 075001.
[60] Allen, Thomas J, Cox, BT, and Beard, Paul C. *Generating photoacoustic signals using high-peak power pulsed laser diodes*. in *Biomedical Optics 2005*. 2005. International Society for Optics and Photonics.
[61] Kohl, A., et al. *An ultra compact laser diode source for integration in a handheld point-of-care photoacoustic scanner*. 2016.
[62] Canal, Celine, et al. *Portable multiwavelength laser diode source for handheld photoacoustic devices*. in *SPIE Photonics Europe*. 2016. SPIE.
[63] Allen, Thomas J and Beard, Paul C. *Dual wavelength laser diode excitation source for 2D photoacoustic imaging*. in *Biomedical Optics (BiOS) 2007*. 2007. International Society for Optics and Photonics.
[64] Rodríguez, Sergio, et al. *Optoacoustic system based on 808-nm high energy short pulse diode laser stacks*. 2017.
[65] Salimian, A, et al. *Laser Diode Induced Lighting Modules*. 2016. Electrochemical Society.
[66] Agano, Toshitaka and Sato, Naoto. *Photoacoustic Imaging System using LED light source*. in *Lasers and Electro-Optics (CLEO), 2016 Conference on*. 2016. IEEE.
[67] Hariri, Ali, et al., *The characterization of an economic and portable LED-based photoacoustic imaging system to facilitate molecular imaging.* Photoacoustics, 2018. **9**: p. 10-20.
[68] Xia, Wenfeng, et al., *Handheld Real-Time LED-Based Photoacoustic and Ultrasound Imaging System for Accurate Visualization of Clinical Metal Needles and Superficial Vasculature to Guide Minimally Invasive Procedures.* Sensors (Basel, Switzerland), 2018. **18**(5).
[69] Allen, Thomas J and Beard, Paul C, *High power visible light emitting diodes as pulsed excitation sources for biomedical photoacoustics.* Biomedical optics express, 2016. **7**(4): p. 1260-1270.
[70] Zhu, Yunhao, et al., *Light Emitting Diodes based Photoacoustic Imaging and Potential Clinical Applications.* Scientific reports, 2018. **8**(1): p. 9885.

[71] Kang, Jeeun, et al. *Toward High-speed Transcranial Photoacoustic Imaging using Compact Near-infrared Pulsed LED Illumination System*. in *SPIE BiOS*. 2016. SPIE.
[72] Wong, Terence TW, et al., *Use of a single xenon flash lamp for photoacoustic computed tomography of multiple-centimeter-thick biological tissue ex vivo and a whole mouse body in vivo.* Journal of biomedical optics, 2016. **22**(4): p. 041003.
[73] Hughes, DW and Barr, JRM, *Laser diode pumped solid state lasers.* Journal of Physics D: Applied Physics, 1992. **25**(4): p. 563.
[74] Hanna, DC and Clarkson, WA, *A review of diode-pumped lasers: advances in lasers and applications.* 1998.
[75] Krupke, William Franklin. *Advanced diode-pumped solid state lasers (DPSSLs): near-term trends and future prospects*. in *High-Power Lasers in Manufacturing*. 2000. International Society for Optics and Photonics.
[76] Wang, Depeng, et al., *Deep tissue photoacoustic computed tomography with a fast and compact laser system.* Biomedical optics express, 2017. **8**(1): p. 112-123.
[77] LeBoulluec, Peter, Liu, Hanli, and Yuan, Baohong, *A cost-efficient frequency-domain photoacoustic imaging system.* American journal of physics, 2013. **81**(9): p. 712-717.
[78] Telenkov, Sergey, et al., *Frequency-domain photoacoustic phased array probe for biomedical imaging applications.* Optics letters, 2011. **36**(23): p. 4560-4562.
[79] Maslov, Konstantin and Wang, Lihong V, *Photoacoustic imaging of biological tissue with intensity-modulated continuous-wave laser.* Journal of biomedical optics, 2008. **13**(2): p. 024006-024006-5.
[80] Daoudi, K., et al. *Handheld probe for portable high frame photoacoustic/ultrasound imaging system*. in *SPIE BiOS*. 2013. SPIE.
[81] Pramanik, Manojit. *Deep imaging with low-cost photoacoustic tomography system with pulsed diode laser*. in *International Conference on Optical and Photonic Engineering (icOPEN2015)*. 2015. SPIE.
[82] Dai, Xianjin, Yang, Hao, and Jiang, Huabei, *In vivo photoacoustic imaging of vasculature with a low-cost miniature light emitting diode excitation.* Optics letters, 2017. **42**(7): p. 1456-1459.
[83] Khan, Asif, et al., *Piezoelectric thin films: an integrated review of transducers and energy harvesting.* Smart Materials and Structures, 2016. **25**(5): p. 053002.
[84] Sappati, Kiran and Bhadra, Sharmistha, *Piezoelectric Polymer and Paper Substrates: A Review.* Sensors, 2018. **18**(11): p. 3605.
[85] Ketterling, Jeffrey A, et al., *Design and fabrication of a 40-MHz annular array transducer.* IEEE transactions on ultrasonics, ferroelectrics, and frequency control, 2005. **52**(4): p. 672-681.
[86] Sherar, MD and Foster, FS, *The design and fabrication of high frequency poly (vinylidene fluoride) transducers.* Ultrasonic imaging, 1989. **11**(2): p. 75-94.
[87] Goldberg, Richard L and Smith, Stephen W, *Multilayer piezoelectric ceramics for two-dimensional array transducers.* IEEE transactions on ultrasonics, ferroelectrics, and frequency control, 1994. **41**(5): p. 761-771.
[88] Martin, K Heath, et al., *Dual-frequency piezoelectric transducers for contrast enhanced ultrasound imaging.* Sensors, 2014. **14**(11): p. 20825-20842.
[89] Muralt, Paul, et al., *Piezoelectric micromachined ultrasonic transducers based on PZT thin films.* IEEE transactions on ultrasonics, ferroelectrics, and frequency control, 2005. **52**(12): p. 2276-2288.

[90] Zipparo, Michael J, Shung, K Kirk, and Shrout, Thomas R, *Piezoceramics for high-frequency (20 to 100 MHz) single-element imaging transducers.* IEEE transactions on ultrasonics, ferroelectrics, and frequency control, 1997. **44**(5): p. 1038-1048.
[91] Li, Changhui, Ku, Geng, and Wang, Lihong V, *Negative lens concept for photoacoustic tomography.* Physical Review E, 2008. **78**(2): p. 021901.
[92] Zhang, Edward, Laufer, Jan, and Beard, Paul, *Backward-mode multiwavelength photoacoustic scanner using a planar Fabry-Perot polymer film ultrasound sensor for high-resolution three-dimensional imaging of biological tissues.* Applied optics, 2008. **47**(4): p. 561-577.
[93] Upputuri, Paul Kumar, Sivasubramanian, Kathyayini, and Pramanik, Manojit. *High speed photoacoustic tomography system with low cost portable pulsed diode laser*. 2015.
[94] Sivasubramanian, Kathyayini and Pramanik, Manojit. *High frame rate photoacoustic imaging using clinical ultrasound system*. 2016.
[95] Sivasubramanian, Kathyayini and Pramanik, Manojit, *High frame rate photoacoustic imaging at 7000 frames per second using clinical ultrasound system.* Biomedical Optics Express, 2016. **7**(2): p. 312-323.
[96] Kalva, Sandeep Kumar, Hui, Zhe Zhi, and Pramanik, Manojit. *Multiple single-element transducer photoacoustic computed tomography system*. in *Photons Plus Ultrasound: Imaging and Sensing 2018*. 2018. International Society for Optics and Photonics.
[97] Upputuri, Paul Kumar, et al. *Pulsed laser diode photoacoustic tomography (PLD-PAT) system for fast in vivo imaging of small animal brain*. in *Photons Plus Ultrasound: Imaging and Sensing 2017*. 2017. International Society for Optics and Photonics.
[98] Zafar, Mohsin, et al., *Development of Low-Cost Fast Photoacoustic Computed Tomography: System Characterization and Phantom Study.* Applied Sciences, 2019. **9**(3): p. 374.
[99] Zafar, Mohsin, et al. *Low-cost fast photoacoustic computed tomography: phantom study*. in *Photons Plus Ultrasound: Imaging and Sensing 2019*. 2019. International Society for Optics and Photonics.
[100] Beard, Paul, *Biomedical photoacoustic imaging.* Interface focus, 2011. **1**(4): p. 602-631.
[101] Haltmeier, Markus, et al., *Thermoacoustic computed tomography with large planar receivers.* Inverse problems, 2004. **20**(5): p. 1663.
[102] Szabo, Thomas L and Lewin, Peter A, *Ultrasound transducer selection in clinical imaging practice.* Journal of Ultrasound in Medicine, 2013. **32**(4): p. 573-582.
[103] Angelsen, Bjørn AJ, et al., *Which transducer array is best?* European Journal of Ultrasound, 1995. **2**(2): p. 151-164.
[104] Daoudi, Khalid, et al., *Handheld probe for portable high frame photoacoustic/ultrasound imaging system*. Vol. 8581. 2013.
[105] Chinni, Bhargava, et al. *Multi-acoustic lens design methodology for a low cost C-scan photoacoustic imaging camera*. in *SPIE BiOS*. 2016. SPIE.
[106] Wang, Xueding, Chamberland, David L, and Jamadar, David A, *Noninvasive photoacoustic tomography of human peripheral joints toward diagnosis of inflammatory arthritis.* Optics letters, 2007. **32**(20): p. 3002-3004.
[107] Li, Mucong, et al., *Linear array-based real-time photoacoustic imaging system with a compact coaxial excitation handheld probe for noninvasive sentinel lymph node mapping.* Biomedical optics express, 2018. **9**(4): p. 1408-1422.

[108] Yin, Bangzheng, et al., *Fast photoacoustic imaging system based on 320-element linear transducer array.* Physics in Medicine & Biology, 2004. **49**(7): p. 1339.
[109] Gao, Fei, et al., *A Prototype for a Palm-sized Photoacoustic Sensing Unit*, in *X-Acoustics: Imaging and Sensing*. 2015.
[110] Mihailo, D. Rabasović, et al., *Low-cost, portable photoacoustic setup for solid samples.* Measurement Science and Technology, 2009. **20**(9): p. 095902.
[111] Starecki, Tomasz and Grajda, Marcin. *Low cost miniature data acquisition and control system for photoacoustic experiments*. in *Photonics Applications in Astronomy, Communications, Industry, and High-Energy Physics Experiments IV*. 2006. SPIE.
[112] Jiang, D., et al., *Low-Cost Photoacoustic Tomography System Based on Multi-Channel Delay-Line Module.* IEEE Transactions on Circuits and Systems II: Express Briefs, 2019. **66**(5): p. 778-782.
[113] Wang, Depeng, Wu, Yun, and Xia, Jun, *Review on photoacoustic imaging of the brain using nanoprobes.* Neurophotonics, 2016. **3**(1): p. 010901.
[114] Xia, Jun and Wang, Lihong V, *Small-animal whole-body photoacoustic tomography: a review.* IEEE Transactions on Biomedical Engineering, 2014. **61**(5): p. 1380-1389.
[115] Zhong, Hongtao, et al., *Review of Low-Cost Photoacoustic Sensing and Imaging Based on Laser Diode and Light-Emitting Diode.* Sensors (Basel, Switzerland), 2018. **18**(7).
[116] Xia, Jun, Yao, Junjie, and Wang, Lihong V, *Photoacoustic tomography: principles and advances.* Electromagnetic waves (Cambridge, Mass.), 2014. **147**: p. 1.
[117] Erfanzadeh, Mohsen and Zhu, Quing, *Photoacoustic imaging with low-cost sources; A review.* Photoacoustics, 2019.
[118] Zafar, Mohsin, et al. *Photoacoustic signal enhancement using a novel adaptive filtering algorithm*. in *Photons Plus Ultrasound: Imaging and Sensing 2019*. 2019. International Society for Optics and Photonics.
[119] Mahmoodkalayeh, Sadreddin, et al., *Low temperature-mediated enhancement of photoacoustic imaging depth.* Scientific reports, 2018. **8**(1): p. 4873.
[120] Mozaffarzadeh, Moein, et al., *Double-Stage Delay Multiply and Sum Beamforming Algorithm: Application to Linear-Array Photoacoustic Imaging.* IEEE Transactions on Biomedical Engineering, 2018. **65**(1): p. 31-42.
[121] Mozaffarzadeh, Moein, et al., *Linear-array photoacoustic imaging using minimum variance-based delay multiply and sum adaptive beamforming algorithm.* Journal of biomedical optics, 2018. **23**(2): p. 026002.
[122] Manwar, Rayyan, et al., *Photoacoustic Signal Enhancement: Towards Utilization of Low Energy Laser Diodes in Real-Time Photoacoustic Imaging.* Sensors, 2018. **18**(10): p. 3498.
[123] Fayyaz, Zahra, et al., *Simulated annealing optimization in wavefront shaping controlled transmission.* Applied optics, 2018. **57**(21): p. 6233-6242.
[124] Paridar, Roya, et al., *Double minimum variance beamforming method to enhance photoacoustic imaging.* arXiv preprint arXiv:1802.03720, 2018.
[125] Gao, Fei, et al., *Single laser pulse generates dual photoacoustic signals for differential contrast photoacoustic imaging.* Scientific reports, 2017. **7**(1): p. 626.
[126] Bell, M. A. L., et al. *Improved contrast in laser-diode-based photoacoustic images with short-lag spatial coherence beamforming*. in *2014 IEEE International Ultrasonics Symposium*. 2014.
[127] Fayyaz, Zahra, et al., *A Comparative Study of Optimization Algorithms for Wavefront Shaping.* Journal of Innovative Optical Health Sciences, 2019.

[128] Meng, Jing, et al., *High-speed, sparse-sampling three-dimensional photoacoustic computed tomography in vivo based on principal component analysis.* Journal of biomedical optics, 2016. **21**(7): p. 076007.
[129] Arridge, Simon, et al., *Accelerated high-resolution photoacoustic tomography via compressed sensing.* Physics in Medicine & Biology, 2016. **61**(24): p. 8908.
[130] Özbek, Ali, Deán-Ben, Xosé Luís, and Razansky, Daniel, *Optoacoustic imaging at kilohertz volumetric frame rates.* Optica, 2018. **5**(7): p. 857-863.
[131] Antholzer, Stephan, Haltmeier, Markus, and Schwab, Johannes, *Deep learning for photoacoustic tomography from sparse data.* Inverse Problems in Science and Engineering, 2019. **27**(7): p. 987-1005.
[132] Stranneby, Dag, *Digital signal processing and applications*. 2004: Elsevier.
[133] Pllana, Sabri and Xhafa, Fatos, *Programming Multicore and Many-core Computing Systems*. Vol. 86. 2017: John Wiley & Sons.
[134] Morimoto, Ryohei, et al., *Randomly polarised beam produced by magnetooptically Q-switched laser.* Scientific reports, 2017. **7**(1): p. 15398.
[135] Qiu, Yongqiang, et al., *Piezoelectric micromachined ultrasound transducer (PMUT) arrays for integrated sensing, actuation and imaging.* Sensors, 2015. **15**(4): p. 8020-8041.
[136] Manwar, Rayyan and Chowdhury, Sazzadur, *Experimental analysis of bisbenzocyclobutene bonded capacitive micromachined ultrasonic transducers.* Sensors, 2016. **16**(7): p. 959.
[137] Manwar, R, et al., *Fabrication and characterization of a high frequency and high coupling coefficient CMUT array.* Microsystem Technologies, 2017. **23**(10): p. 4965-4977.
[138] Miida, Yusuke and Matsuura, Yuji, *All-optical photoacoustic imaging system using fiber ultrasound probe and hollow optical fiber bundle.* Optics express, 2013. **21**(19): p. 22023-22033.
[139] Johnson, Jami L, van Wijk, Kasper, and Merrilees, Mervyn. *All-optical Photoacoustic and laser-ultrasound imaging of fixed arterial tissue (Conference Presentation)*. in *Photons Plus Ultrasound: Imaging and Sensing 2017*. 2017. International Society for Optics and Photonics.
[140] Bauer-Marschallinger, Johannes, Felbermayer, Karoline, and Berer, Thomas, *All-optical photoacoustic projection imaging.* Biomedical optics express, 2017. **8**(9): p. 3938-3951.